# Status of CSNS H- ion source

LIU Shengjin(刘盛进)[1,2;1)], HUANG Tao(黄涛)[1,2], OUYANG Huafu(欧阳华甫)[1,2], ZHAO Fuxiang(赵富祥)[1,2], XIAO Yongchuan(肖永川)[1,2], LV Yongjia(吕永佳)[1,2], CAO Xiuxia(曹秀霞)[1,2], XUE Kangjia(薛康佳)[1,2], ZHANG Junsong(张俊嵩)[1,2], XU Taoguang(徐韬光)[1,2], LI Fang(李芳)[1,2], LU Yanhua(卢艳华)[1,2], LI Gang(李刚)[1,2], YANG Lei(杨雷)[3], LI Yi[3](李仪)

1 China Spallation Neutron Source (CSNS), Institute of High Energy Physics (IHEP), Chinese Academy of Sciences (CAS)，Dongguan 523803, China

2 Dongguan Institute of Neutron Science (DINS), Dongguan 523808, China

3, Dongguan University of Technology, Dongguan 523808, China

**Abstract:** A new H- ion source has been installed successfully and will be used to serve the China Spallation Neutron Source (CSNS). In this paper, we report various components of the ion source, including discharge chamber, temperature，cooling system, extraction electrodes, analyzing magnet, remote control system and so on. Compared to the previous experimental ion source, some improvements have been made to make the ion source more compact and convenient. In the present arrangement, the Penning field is generated by a pair of pole tip extensions on the 90°analyzing magnet instead of by a separate circuit. For the remote control system, F3RP61-2L is applied to the accelerator online control system for the first time. In the running of the ion source, a stable pulse H- beam with a current of 50 mA at an energy of 50 keV is produced. The extraction frequency and pulse width is 25 Hz and 500 μs, respectively. Furthermore, an emittance scanner has been installed and measurements are in progress.

## 1. Instruction

The China Spallation Neutron Source (CSNS) is a multipurpose facility with a 100 kW class proton beam power [1]. In the accelerator design, an H- ion beam with a peak current of 20 mA and a pulse width of 500 μs is accelerated up to 81 MeV by the linac then injected into the RCS at a repetition rate of 25 Hz. The linear accelerator consists of a 50 keV H- Penning surface plasma ion source, a low energy beam transport line, a 3.5 MeV Radio Frequency Quadrupole (RFQ) accelerator, a medium energy beam transport line, an 81 MeV Drift Tube Linear (DTL) accelerator and a high energy beam transport line. In the experimental phase of the ion source, the H- ion source successfully produced an H- ion current of 50 mA with a duty factor of 1.25% (pulse length and repetition rate are 500 μs and 25 Hz, respectively), which satisfies the requirements of the CSNS. Currently, a new ion source has been successfully completed, which will serve the CSNS directly.

## 2. Ion source construction

The basic design of the CSNS ion source has previously been described in [2]. Here, we introduce it briefly. The type of ion source is a Penning surface plasma source [3]. A cross section view of the present CSNS ion source is shown in Figure 1. The ion source consists of a discharge chamber, an extraction electrode, an analyzing magnet, a -50 kV isolation ceramic insulator, an accelerating electrode and a large vacuum chamber with two turbo molecular pumps (TMP). The chamber vacuum is maintained below $5.0 \times 10^{-5}$ Pa. Cesium is used to increase the H- ion production efficiency.

---

1) Email: liusj@ihep.ac.cn    Address: No.1 Zhongziyuan Road, Dalang, Dongguan, China

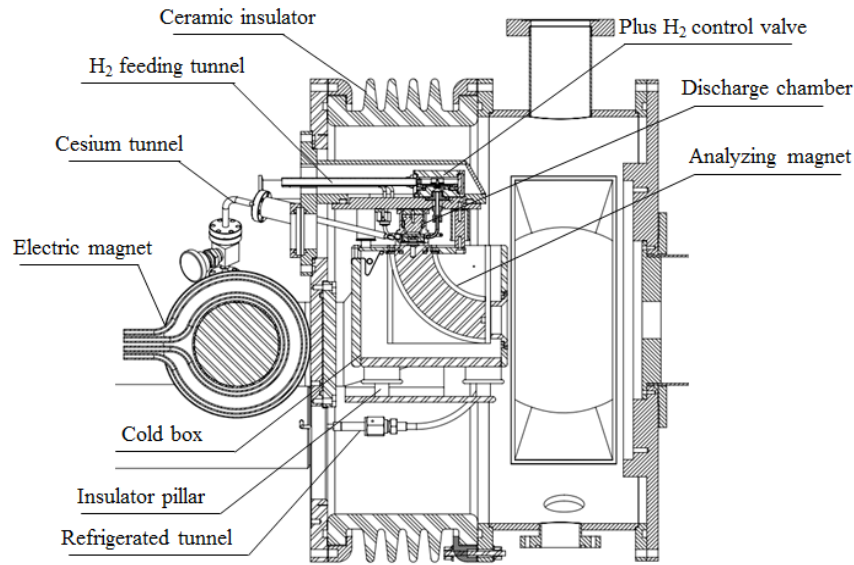

Figure 1 Cross section of CSNS ion source

The discharge chamber comprises a molybdenum anode, cathode and an aperture plate, where the plasma arc is confined. A transverse magnetic Penning field is applied across the discharge chamber. Hydrogen and cesium are fed into the discharge chamber via holes in the anode, which can be more clearly seen in Figure 2. The anode and cathode are housed in the stainless steel source body. The anode is thermally and electrically connected to the source body, whereas the cathode is isolated from the source body using a ceramic spacer. The whole assembly is mounted to the ion source flange, separated by a thin layer of mica to provide electrical isolation for the cathode. The negative ions are extracted through the slit in the aperture plate.

To maintain stable operation of the ion source, the discharge chamber temperatures are an important factor when considering the performance of the ion source, since the discharge electrode surface plays an important role in the plasma physics of the arc. There is a an optimum range (400-600°C) of electrode surface temperature for the production of $H^-$ ions, therefore stable electrode temperature is essential for reliable ion source operation and consistent output currents. Too high an operational temperature will influence the lifetime of the discharge chamber. In normal operation the temperatures of cathode, anode and source body are monitored using a thermocouple. Ion source cooling is provided by air and water cooling. Water is used to cool the mounting flange plate, hydrogen feed system and analyzing magnet, whereas air is used to cool the source body via two pipes in the source body. The temperature of the source body is stabilized via a closed loop control system with air cooling.

The beam extractor consists of 3 electrodes: an extraction electrode, a protection electrode, and an acceleration electrode. For CSNS a platform voltage of -50 kV is required. The protection electrode is installed in front of the acceleration electrode at the low voltage side of the accelerating gap to limit the current to the power supplies in the event of flashover, thus protecting the whole acceleration circuit.

The analyzing magnet is a 90°C sector magnet used to separate the $H^-$ ions from the other negative ion species which are also extracted from the source. It is noted that the analyzing magnet is mounted into a refrigerated coldbox to allow the coldbox to trap cesium vapor escaping from the source.

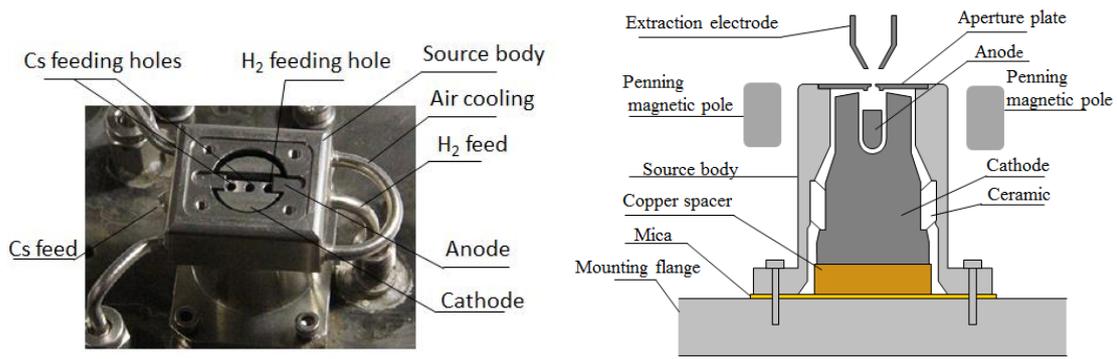

Figure 2 The components of the discharge chamber. The left-hand picture shows the cathode, anode and source body. The cross section of the chamber is shown in the right-hand figure. [4].

In the experimental phase, the Penning field was supplied by a separate magnetic circuit. This was convenient for separate control of the Penning field and the ion source. However, it is found that a weak Penning field is sufficient to maintain pulse discharge of the ion source. In the present arrangement, the separate magnetic circuit is removed and the Penning field is generated by a set of replaceable pole tip extensions on the top of the analyzing magnet. Therefore, the analyzing magnet has another function, to supply the Penning field for the discharge chamber. Since the separate circuit was removed, the first chamber to set the discharge chamber was also removed. The discharge chamber is thus set in the -50kV isolation insulator chamber directly, where ceramic insulator replaces the previous nylon insulator. Figure 3 shows the differences between the experimental phase and the present arrangement of the ion source. The position of the magnet pole pieces is shown in Figure 4.

As it is supplied by the pole tip extensions on top of the analyzing magnet, the Penning field cannot be varied easily. The test results demonstrate that the Penning field supplied from the pole tip extensions satisfies the operational requirements of the ion source. In the process of pulse discharge, we tried to reduce the current of the analyzing magnet to 8 A, and found that the extension Penning field is enough to maintain a stable pulse discharge current. For the ion source, the power supplies mainly include the power supply of the $H_2$ feeding system controlling the $H_2$ flux of, DC power supply, pulse discharge power supply, extraction electrode power supply, analyzing magnet power supply, and cesium oven and transport line power supply. All of them are set on the negative high voltage (-50kV) platform, which is isolated from ground via polysulfone insulated pillars.

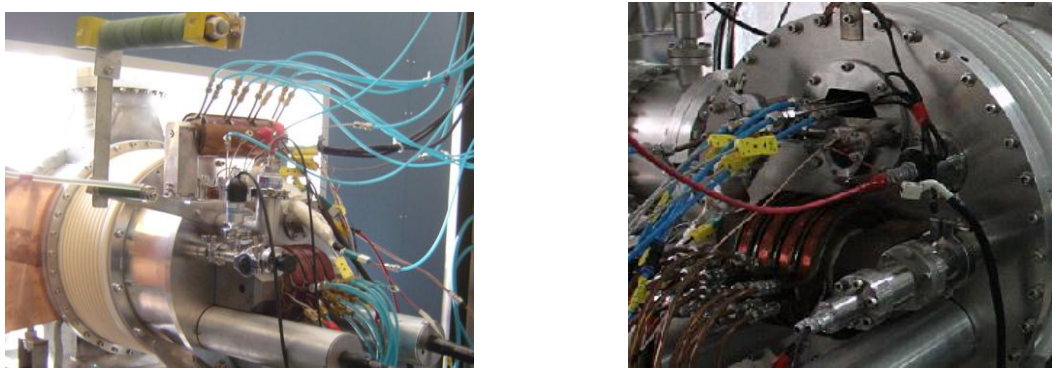

Figure 3 Left picture(experimental phase): the Penning field was provided by a separate circuit. Right picture (present arrangement): the Penning field is provided by the pole tip extension from analyzing magnet, and the separate circuit and first chamber are removed. The discharge chamber is set directly into ceramic insulator chamber.

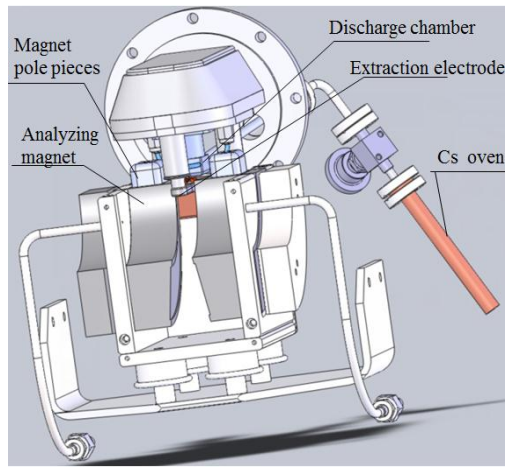

Figure 4 The pole tip extension on the top of the analyzing magnet.

The control system of the ion source consists of a local and remote control system linked via Ethernet. The control system was developed by utilizing Programmable Logic Controllers (PLC) and Experimental Physics and Industrial Control System (EPICS). TCP/IP-based Ethernet is used for communication between the PLC and the PC, which gives a high speed network for transmitting critical signals for the ion source system and provides real time control and monitoring. The EPICS software is adopted to realize the integration of extendable systems and give a user-friendly operator interface. Furthermore, EPICS allows the user to implement the control and monitoring, and create a state notation programs, through a graphical user interface. A closed loop control system is used for temperature and air cooling.

It is noted that F3RP61-2L, which is an embedded computer with a PowerPC CPU, and a platform for the Linux embedded operating system, is applied to an accelerator online control system for the first time in China. The introduction of F3RP61-2L offers a new platform option for EPICS IOC in the CSNS control system, simplifying the standard three-level control system structure into two levels. Since F3RP61-2L synchronously communicates with the PLC device interface through PLC-Bus instead of by the previous Ethernet communication, the response speed of the system is improved. More details of the control system, including soft implementation and PLC interlocks, can be found in [5].

## 3. Operation of the ion source

The operation of the ion source starts from the DC discharge. Normally, the DC discharge lasts for 30 - 40 minutes. In this process, we try to pre-heat the discharge chamber up to the operating temperature of around 450 $^o$C. At the same time, this allows examination of the performance of the ion source including: $H_2$ and cesium deliver systems, extraction pulse power supply, analyzing magnet, vacuum system and so on. Cesium vapor is delivered into the discharge chamber by heating up the cesium oven and transport line at the same time. In the AC power supply design, the highest input voltage is limited to 300V. Therefore, when the DC current reaches several amperes at the discharge voltage of around 300V, the low current DC discharge can be directly switched into the high current pulsed discharge. For the timing of the ion source, the current of the arc discharge pulse is run at 25 Hz with 50 A, 600 μs. The timing of the pulse $H_2$ gas is thus moved up to around 700 μs with a width of 200 μs, while the timing of the extraction voltage pulse is enclosed within the arc discharge pulse and delayed around 100 μs with a width of 500 μs. With an extraction voltage of 17 kV, the intensity of the extraction beam, mainly electrons and H⁻ beam, is around 300 mA. After extraction the beam is bent through the analyzing magnet and further accelerated by a post-acceleration voltage of 33 kV to 50 keV. Finally, a pulse H⁻ beam of 50 mA, 500 μs, 25 Hz is obtained through the ACCT monitor. The operating parameters are shown in Table 1. Figure 5-a shows the various pulses, including $H_2$ gas, arc discharge and extraction voltage. An H⁻ beam pulse is shown in Figure 5-b.

Table 1 Operating parameters of the CSNS H⁻ ion source

| Output energy | 50 keV | Temperature of Cs oven | 150~170 °C |
|---|---|---|---|
| Repetition rate | 25 Hz | Temperature of Cs transport line | ~300 °C |
| Pulse H⁻ beam width | 500 μs | Extraction voltage | 17 kV |
| Pulse H⁻ beam current | 50 mA | Current of Analyzing magnet | 10.7A |
| Flux of $H_2$ | ~10SCCM | Pulse arc current | 50A |
| Pulse arc width | 600 μs | Chamber vacuum | ~3.0×10⁻³ Pa |

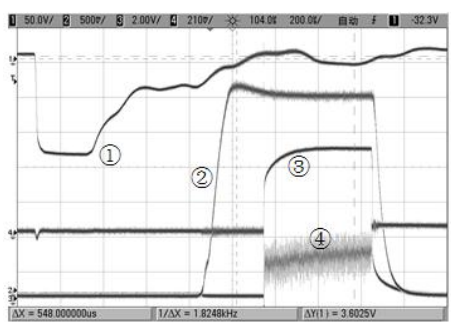 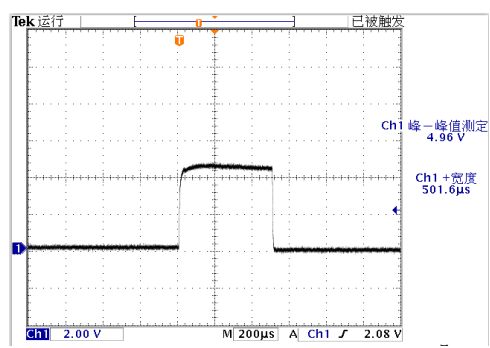

Figure 5 Oscilloscope images of pulse discharge (left Fig.5-a) and H⁻ beam at ACCT (right Fig.5-b). The repetition rate is 25Hz. Fig.5-a ①: pulse hydrogen feeding, ②: pulse discharge current: 50A, ③: pulse extraction voltage: 17kV, ④: extraction current: 240mA (including electrons and H⁻). Fig.5-b: Current of H⁻, 50mA, 500 μs and 25 Hz.

After each beam run, the ion source is overhauled, cleaned and inspected for any damage to its internal elements, with repairs done if necessary. The discharge chamber components including cathode, anode and source body are replaced at each maintenance. Generally, 5 grams of cesium can last for one month with a 150~170 °C cesium oven. The extraction electrode and aperture plate are reused by polishing their surface ultrasonically. Figure 6 shows a photograph of the cathode after approximately one week of operation. The surface erosion on the both cathode surfaces is obvious; this is due to bombarding by a large number of particles. We also find that the anode surface is covered with Mo filings sputtered from the cathode.

Surface erosion is probably the most important factor limiting an electrode's useful lifetime. The anode and cathode surfaces are both eroded, and the erosion rate depends on the electrode material, the uniformity of the Penning discharge and the source operating condition. A lower DC charge time is beneficial for the cathode duration. Furthermore, the purity and machining techniques used for the Mo are important for the cathode and anode. Pure Mo metal is desirable to increase the lifetime of the discharge chamber.

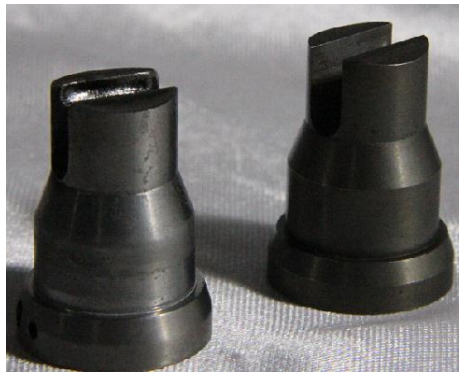

Figure 6 Used cathode (left) and unused cathode (right). For the used cathode, the surface erosion is obvious on its surfaces.

## 4. Emittance measurement

Bam diagnosis is being performed as follows. An emittance scanner is directly located behind the solenoid. Preliminary measurements of the ion source emittance have been made. In both x and y planes, a slit-slit emittance scanner is used. Secondary electron suppression is provided. Here, the beam energy was kept constant at 50 keV. The source provided an H⁻ current of 53 mA at X and Y plane. The emittance plots are shown in Figure 7. The horizontal and vertical emittances were around 0.892 and 0.742πmmmrad, respectively. Furthermore, the relation between emittance and current is deduced via the emittance phase space, which is plotted in Figure 8. The current decreases rapidly with decreasing emittance. At 0.2πmmmrad, the corresponding current is around 15 mA and 25 mA in the x and y planes, respectively. The basic requirements for the RFQ entrance emittance are met.

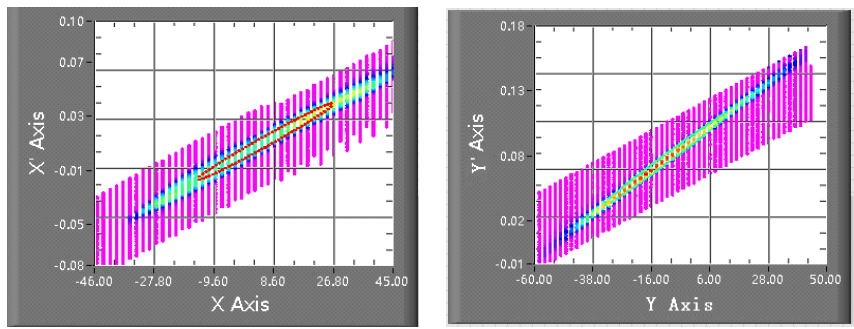

Figure 7 Horizontal and vertical emittance for a 50 keV H⁻ beam. X-plane: 0.892πmmmrad at the H⁻ current of 53mA, Y-plane: 0.742πmmmrad at the H⁻ current of 53 mA. (Color online)

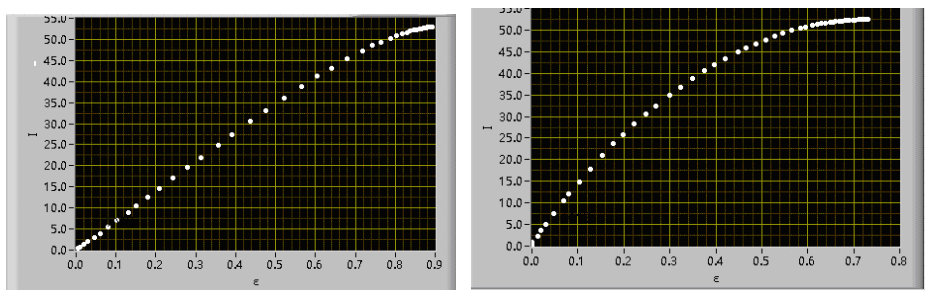

Figure 8 Relation between emittance and beam current. Left: X plane. Right: Y plane. At 0.2πmmmrad, current of X plane and Y plane is 15 mA and 25 mA, respectively.(Color online)

## 5. Summary and future work

The CSNS Penning type ion source has been completed. All parameters have been optimized for the current source design and a pulse H⁻ beam with 25 Hz, 500 μs, 50mA is obtained. At present, the ion source will be applied to CSNS and supply a stable H⁻ beam. Compared to the previous experimental ion source, a significant improvement is that the pole tip extension on the top of the analyzing magnet is applied to the ion source as the across Penning field. In view of the structure and dismounting of the ion source, such a design is convenient for the mounting of the discharge chamber. The results presented prove that the design of the extension Penning field is reliable and strong enough to maintain the pulse discharge and produce a stable H⁻ beam. Furthermore, preliminary emittance measurements have been taken and further emittance tests are now being carried out, which will reveal more information about the H⁻ beam. The ion source performance will be optimized further depending on the results of these measurements. In the future, LEBT and RFQ installations will be prepared and a front-end platform for the CSNS constructed.

### Acknowledgements

We would like to thank Dongguan University of Technology for their help, and the CSNS members of the accelerator center for their contribution during the installation of the ion source.